\documentclass[prl,aps,twocolumn,superscriptaddress]{revtex4-2}
\usepackage{graphicx,color}
\usepackage{amsthm}
\usepackage{amsfonts}
\usepackage{algorithmic}
\usepackage{enumerate}
\usepackage{latexsym}
\usepackage{amsmath}
\usepackage{amssymb}
\usepackage{subfig}
\usepackage{adjustbox}

\usepackage[colorlinks=true,citecolor=blue,linkcolor=blue]{hyperref}
\def\avg#1{\left\langle#1\right\rangle}

\usepackage{dcolumn}
\usepackage{bm}

\usepackage{caption}
\usepackage[T1]{fontenc}
\usepackage{mathptmx}

\begin{document} 
\title{Magnetic correlations and superconducting pairing near higher-order Van Hove singularities}
\author{Zheng Wei}
\affiliation{School of Physics and Astronomy, and Key Laboratory of Multiscale Spin Physics (Ministry of Education), Beijing Normal University, Beijing 100875, China\\}

\author{Yanmei Cai}
\affiliation{School of Physics and Astronomy, and Key Laboratory of Multiscale Spin Physics (Ministry of Education), Beijing Normal University, Beijing 100875, China\\}

\author{Boyang Wen}
\affiliation{School of Physics and Astronomy, and Key Laboratory of Multiscale Spin Physics (Ministry of Education), Beijing Normal University, Beijing 100875, China\\}

\author{Tianxing Ma}
\email{txma@bnu.edu.cn}
\affiliation{School of Physics and Astronomy, and Key Laboratory of Multiscale Spin Physics (Ministry of Education), Beijing Normal University, Beijing 100875, China\\}

\begin{abstract}
{Higher-order Van Hove singularities in strongly correlated electron systems provide a fertile ground for emergent electronic orders and superconductivity. This study investigates the interplay between magnetic fluctuations and superconducting pairing near higher-order Van Hove singularities on the honeycomb lattice, a paradigmatic platform relevant to graphene. By incorporating third-nearest-neighbor hopping \(t''\), we uncover a universal crossover: ferromagnetic fluctuations dominate below the higher-order Van Hove filling, while antiferromagnetic fluctuations take over toward half filling. A key finding is that the already dominant \(f_n\)-wave pairing is enhanced in the critical region of this magnetic crossover by the higher-order Van Hove. This enhancement is driven by the synergistic effect of the higher-order Van Hove singularities-induced divergent density of states and the competing magnetic fluctuations. Although increased hopping parameters generally suppress superconducting correlation, we identify a critical  \(t''\) that anomalously enhances pairing via the higher-order Van Hove renormalization. Furthermore, the nearest-neighbor Coulomb interaction suppresses the pairing correlation function in a sign-independent manner. Our results clarify the competitive mechanisms between magnetic fluctuations and unconventional superconductivity in higher-order Van Hove singularities systems, offering a theoretical basis for tailoring quantum phases in graphene-based materials via band engineering.
}\end{abstract}

\date{Version 16.0 -- \today}

\maketitle
\renewcommand{\thesection}{\Roman{section}}
\section{Introduction}

In the realm of strongly correlated electron systems, the exploration of unconventional superconductivity stands as one of the most captivating frontiers in condensed matter physics\cite{RevModPhys.84.1383}. Unlike conventional superconductors, where phonon-mediated pairing gives rise to Bardeen-Cooper-Schrieffer (BCS) superconductivity, unconventional superconductors typically emerge from intricate electronic interactions, yielding exotic pairing symmetries and non-trivial order parameters\cite{RevModPhys.84.1383}. Extensive investigations have demonstrated that magnetic fluctuations driven by strong electron correlations play a pivotal role in mediating Cooper pairing, as seen in doped cuprates\cite{BF01303701}, iron-based superconductors\cite{ja800073m}, and heavy-fermion systems\cite{PhysRevLett.43.1892}. Specifically, triplet pairing may arise from ferromagnetic-like spin fluctuations, while singlet pairing with $d$-wave symmetry is often associated with antiferromagnetic spin fluctuations\cite{RevModPhys.84.1383,PhysRevB.90.245114}. This dichotomy undergoes the ongoing debate about the competition between different superconducting pairing symmetries and magnetic modes.  

The influence of electronic correlations on superconductivity can be significantly amplified in systems with singular features in their electronic density of states (DOS). Materials hosting Van Hove singularities (VHS) have drawn substantial attention due to the interplay between magnetism and superconductivity, where VHS-enhanced structural and magnetic instabilities may compete with superconducting states\cite{PhysRevB.107.184504,PhysRevB.111.075114,PhysRevB.109.155118,PhysRevLett.131.026601,PhysRevB.107.115105}. A striking example is the magic-angle graphene superlattice system, in which scientists observed unconventional superconductivity closely tied to strong electron correlation effects\cite{Cao2018B,PhysRevB.107.245102,PhysRevResearch.4.L012013}. This correlation is further amplified by the extreme DOS induced by VHS, offering critical insights into the understanding of unconventional superconductivity\cite{nature26154}.  

In recent years, an especially intriguing scenario has emerged in materials with higher-order Van Hove (HOVH) singularities, where the electronic DOS diverges in a power-law manner due to the presence of saddle points with higher-order band flattening\cite{s41467-019-13670-9,PhysRevB.107.184504,s41467-022-29828-x,PhysRevResearch.6.043132,PhysRevResearch.5.L042006}. Unlike conventional VHS, which exhibit a logarithmic DOS divergence, HOVH systems can sustain markedly enhanced electronic correlations, amplifying the effects of interactions such as spin fluctuations, charge instabilities, and unconventional superconductivity\cite{PhysRevLett.99.070401,PhysRevB.85.235408}.  

Theoretical and experimental evidence is emerging to suggest that strong electron-electron correlation effects, which are greatly amplified by the extreme DOS induced by HOVH, may give rise to a distinct type of chiral superconductivity, characterized by a unique pairing symmetry that deviates from conventional superconducting mechanisms\cite{nphys2208}. These singularities have also been shown to trigger quantum phase transitions, leading to substantial modifications in the topological properties of materials and even the emergence of topological superconductivity\cite{PhysRevResearch.4.L012013,HAN2024319}. The HOVH not only enhances the electronic anisotropy of materials, instigating a competition between ferromagnetic (FM) and antiferromagnetic (AFM) fluctuations\cite{PhysRevB.107.184504}, but also promotes Stoner-type magnetism due to the extreme DOS. Moreover, it provides a favorable environment for long-range electron pairing, potentially stabilizing superconducting phases beyond the standard BCS paradigm\cite{RevModPhys.84.1383}.  

Inspired by the intriguing correlated phenomena discovered in graphene-based systems, the Hubbard model on a honeycomb lattice has become a paradigmatic theoretical framework for exploring strong electron correlations in quantum materials. Beyond its Dirac physics at half filling, this prototypical system exhibits VHS in its low-energy spectrum, with the saddle points of the band structure leading to a divergent DOS\cite{10.1063/1.3485059}. Notably, when these saddle points acquire higher-order characteristics (e.g., through strain, twisting, or doping), the singularities evolve into HOVH, leading to a more dramatic amplification of the DOS. In such regimes, the enhanced electronic correlations near HOVH create a fertile platform for intertwined orders, including unconventional superconductivity and complex magnetic phases\cite{JIA2022128175}.

While HOVH can be engineered in various lattices such as square, triangular, and kagome\cite{10.1038/s41467-024-53650-2, HAN2024319, 10.1038/s41467-024-51618-w,PhysRevB.107.184504}, the honeycomb lattice is a paradigmatic platform due to its direct relevance to graphene and the clear, controllable pathway to HOVH via further-neighbor hoppings\cite{PhysRevB.102.125141}. This work leverages these advantages to systematically unravel the interplay between HOVH, magnetic fluctuations, and pairing symmetry—a complex nexus that remains less explored across lattice geometries.

The Hubbard model on a honeycomb lattice has been intensively studied for decades. At half filling, quantum Monte Carlo (QMC) studies indicate that when the on-site Coulomb repulsion energy \(U\) exceeds the critical value \(U_c\approx3.8|t|\), where \(t\) is the nearest-neighbor (NN) hopping, the system undergoes a transition from a semimetal to a Mott insulator, accompanied by the emergence of AFM long-range order\cite{PhysRevLett.120.116601, srep00992, PhysRevX.3.031010}. When the system is away from half filling, electron correlations may induce unconventional superconductivity: near half filling, superconducting pairing is often dominated by the \(d + id\) wave (chiral \(d\) wave) \cite{PhysRevB.84.121410, JIA2022128175, Fang_2021, PhysRevB.100.115135,PhysRevB.104.035104, PhysRevB.107.245106, PhysRevB.110.085103, s11467-022-1236-4, Li_2022,PhysRevB.94.115105}, which originates from the mediation of AFM spin fluctuations. In the large-doping region (\(\avg{n}\simeq0.2\)), when the Fermi level is close to the HOVH singularity at the bottom of the energy band induced by the next-nearest-neighbor (NNN) hopping term \(t'\), the \(p+ip\) wave (chiral \(p\) wave) triplet pairing becomes dominant\cite{PhysRevB.90.245114, PhysRevB.102.125125, PhysRevB.92.035132, PhysRevB.92.174503}. This dominant superconducting pairing is driven by FM fluctuations, and the power-law divergence of the DOS near the HOVH may significantly increase the superconducting critical temperature. However, although the role of HOVH in enhancing electron correlations and superconducting pairing has been preliminarily explored, the reported filling region is rather low, making experimental detection nearly impossible\cite{PhysRevB.90.245114}.  

In honeycomb lattices, the complex interplay between HOVH singularities and correlation-driven instabilities—such as superconductivity and magnetic order—remains a profound mystery. This work expands the existing framework by incorporating the third-nearest-neighbor (TNN) hopping parameter \(t''\): using the determinant quantum Monte Carlo (DQMC) method, we investigate the impact of \(t''\) on magnetic correlations and observe a crossover between FM and AFM fluctuations at different electron fillings. By employing constrained-path Monte Carlo (CPMC) method, we focus on electron correlation-driven superconducting pairing near HOVH under various conditions of electron filling \(\langle n\rangle\), hopping parameters \(t'\) and \(t''\), and nearest-neighbor Coulomb interaction \(V\). Our findings reveal that near the HOVH filling, superconducting pairing with $f_n$ symmetry becomes dominant, and this pairing may be driven by competing FM/AFM fluctuations.

\section{Model and Methods}
\begin{figure}[ht]
	\centering
	\includegraphics[scale=0.26]{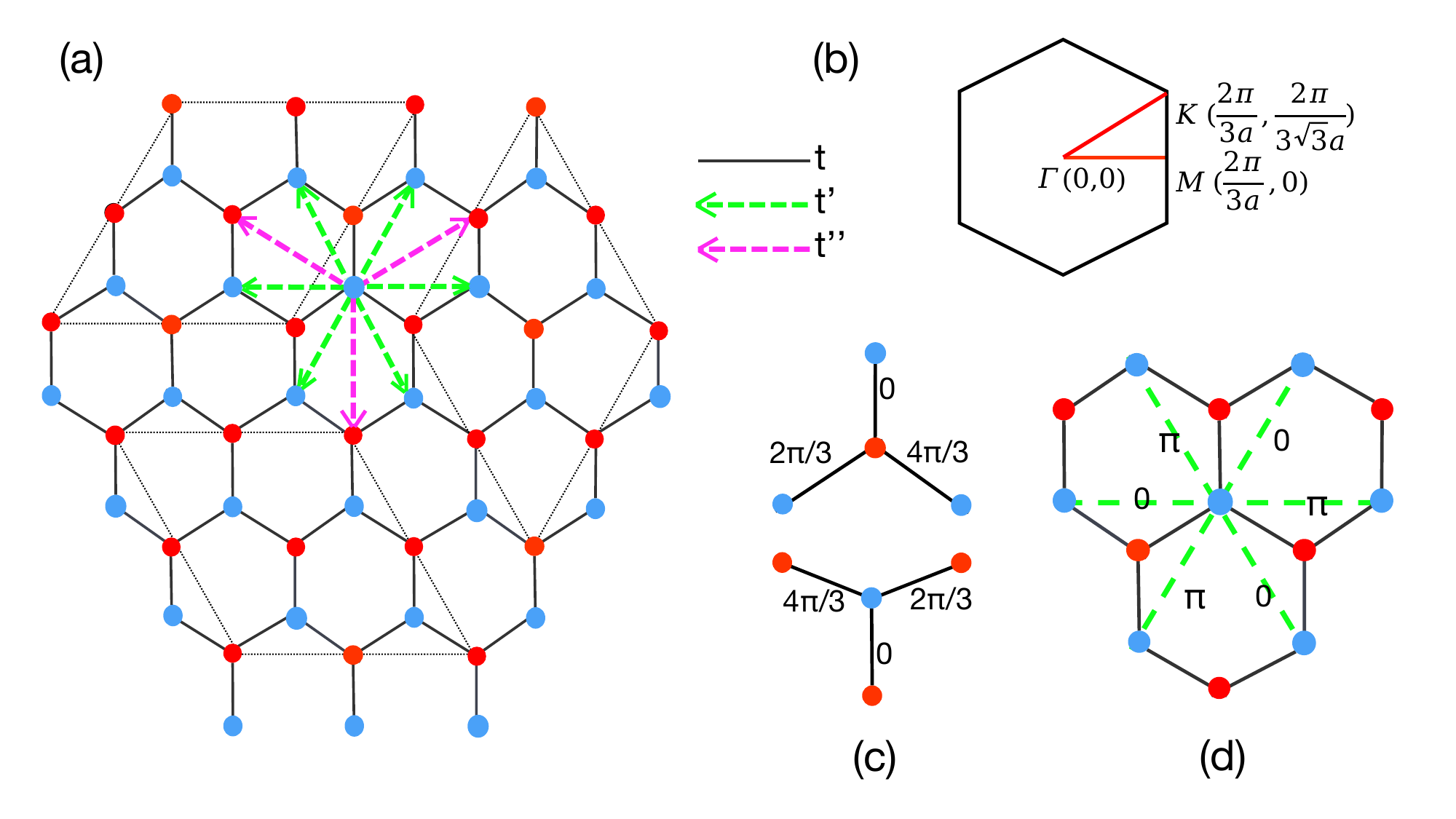}
	\caption{(a) Sketch of the honeycomb lattice with double-27 sites of $N=2\times3L^2$ with $L=3$. (b) The first Brillouin zone and high symmetry points. (c) The form of NN \( d+id \) pairing symmetry. (d) The form of NNN \( f_n \) pairing symmetry.}
	\label{Fig:sketch}
\end{figure}

The system we simulated is sketched in Fig.~\ref{Fig:sketch}(a), and the honeycomb lattice structure consists of two sublattices: the \(A\) sublattice, represented by blue circles, and the \(B\) sublattice, denoted by red circles. The first Brillouin zone, depicted in Fig.~\ref{Fig:sketch}(b), shows high-symmetry points along standard paths, essential for electronic structure analysis.

We began by constructing a theoretical model for the honeycomb lattice based on the Hubbard model:
\begin{align}
H&=H_{k}+H_{U}-\mu \sum_{i\sigma} n_{i\sigma} \nonumber,\\
H_k&=-t\sum_{\langle i,j \rangle \sigma}c_{i\sigma}^\dagger c_{j\sigma}-t'\sum_{\langle\langle i,j\rangle\rangle \sigma}c_{i\sigma}^\dagger c_{j\sigma} \nonumber\\
&-t''\sum_{\langle\langle\langle i,j \rangle\rangle\rangle \sigma}c_{i\sigma}^\dagger c_{j\sigma}+\text{H.c.} \nonumber,\\
H_U&=U \sum_i n_{i\uparrow} n_{i\downarrow}+V \sum_{\langle i,j \rangle \sigma \sigma'}n_{i\sigma}n_{j\sigma'}.
\label{eq:model}
\end{align}
Here, \(t\), \(t'\), and \(t''\) denote the hopping integrals between nearest-neighbor, next-nearest-neighbor, and third-nearest-neighbor sites, respectively. The indices \(i\) and \(j\) signify lattice sites. The operators \(c_{i\sigma}^\dagger\) and \(c_{j\sigma}\) are used to create and annihilate an electron with spin \(\sigma\) at sites \(i\) and \(j\), respectively. The operator \(n_{i\sigma}=c_{i\sigma}^\dagger c_{i\sigma}\) quantifies the number of electrons with spin \(\sigma\) at site \(i\). \(U\) represents the on-site Coulomb interaction strength, while \(V\) stands for the NN Coulomb interaction strength. The chemical potential \(\mu\) is utilized to regulate the electron concentration.

In our calculations, the hopping parameters \(t'\), and \(t''\) are expressed in units of the nearest-neighbor hopping \(t\). For graphene, \(t\) is typically \(\approx 2.0\ \text{eV}\)\cite{PhysRevB.90.245114}, and  the next-nearest-neighbor hopping \(t'\) is estimated to range from \(0.10t\) to \(0.30t\) in relevant systems\cite{PhysRevB.88.165427,PhysRevB.76.081406}. The critical relation in Eq. (2) dictates the emergence of the HOVH. For a representative \(t' = 0.20t\), it yields a critical \(t'' = 0.15t\), which is the value at which we observe the anomalous enhancement of \(f_n\) wave pairing. 

Compared to the pristine honeycomb lattice with only NN hopping $t$, the inclusion of long-range hoppings $t'$ and $t''$ enhances electron delocalization while intensifying geometric frustration. This dual effect modifies the electronic band structure and shifts the HOVH singularity position. To quantitatively determine the HOVH location, we analyze the noninteracting ($U=0$) band dispersion:
\[ E_ \pm ( \vec{k} ) = \pm  \mid t\alpha( \vec{k} ) +t'' \gamma ( \vec{k} )\mid  -t' \beta ( \vec{k} ). \]
Here $ \alpha( \vec{k} )=\sum\limits_{i=1}^3 e^{-i \vec{k} \cdot  \vec{\mathbf{a}}_i  }$, $\beta( \vec{k} ) = \sum\limits_{i=1}^6 e^{-i \vec{k} \cdot  \vec{\mathbf{b}}_i }$, and $ \gamma ( \vec{k} ) = \sum\limits_{i=1}^3 e^{-i \vec{k} \cdot  \vec{\mathbf{c}}_i  }$, where $\vec{\mathbf{a}}_i$, $\vec{\mathbf{b}}_i$, and $\vec{\mathbf{c}}_i$ denote NN, NNN, and TNN displacement vectors, respectively. Expanding $E(\vec{k})$ around the $M$ point up to quadratic order, we identify the HOVH condition when the effective mass tensor becomes singular. Specifically, the critical relation\cite{PhysRevB.102.125141} 
\begin{equation}
t'' = \frac{t - 2t'}{4}.
\end{equation}

\begin{figure}[h]
	\centering
	\adjustbox{trim=120 10 138 6, clip}{
	\includegraphics[width=0.92\textwidth]{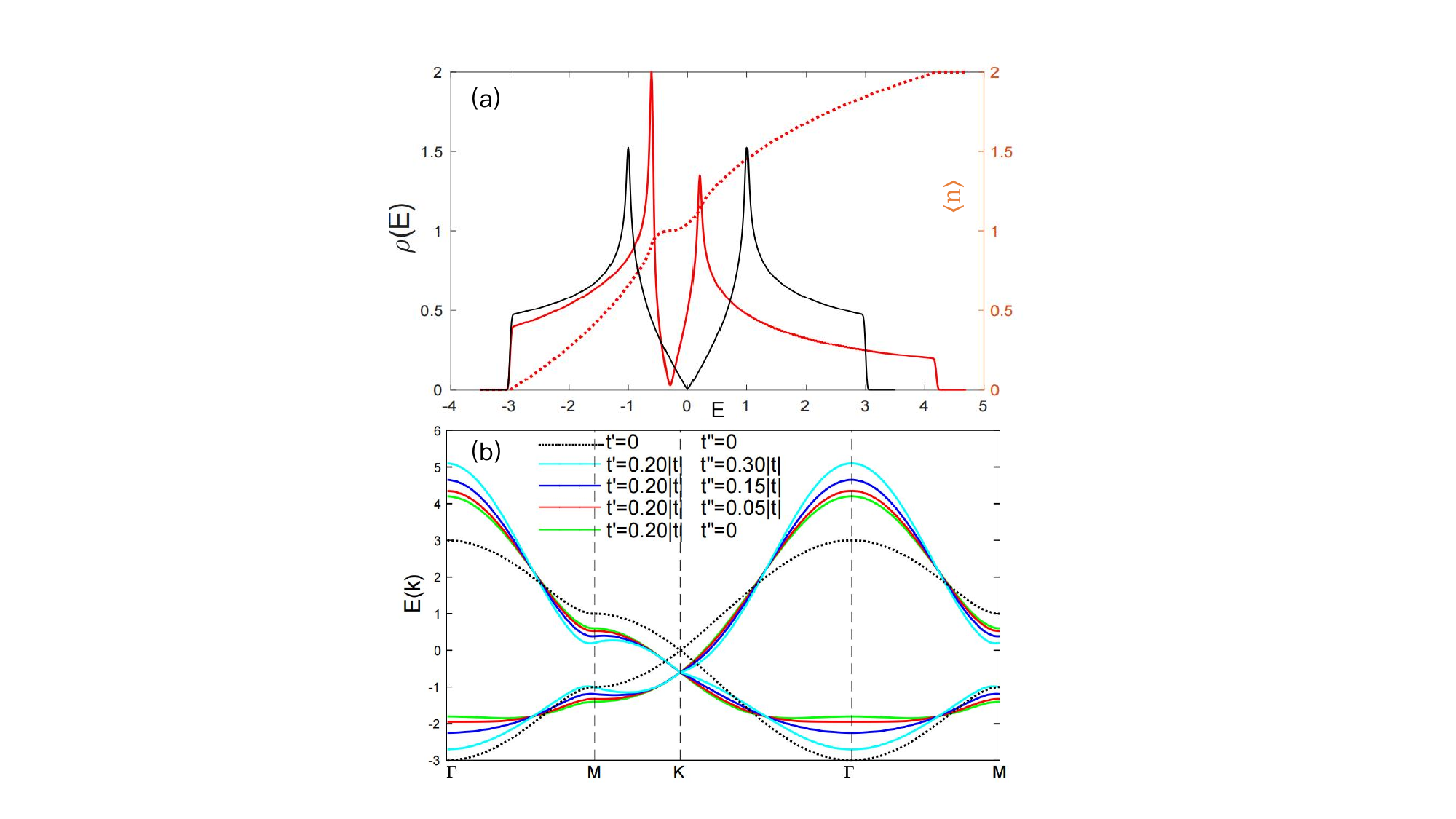}
	}
	\caption{(a) The DOS as functions of energy with \( t = 1.00 \), \( t' = 0\), \( t'' = 0 \) (black solid line) and \( t = 1.00 \), \( t' = 0.10 \), \( t'' = 0.20 \) (red solid line). The red dotted line shows fillings \( \langle n \rangle \) as a function of energy with \( t = 1.00 \), \( t' = 0.10 \), and \( t'' = 0.20 \).
	(b) Noninteracting band dispersions for different \( t'' \) values at fixed \( t=1.0, t'=0.20 \)}
	\label{fig:2}
\end{figure}

This critical relation signals a topological transition at the saddle point, where the conventional VHS evolves into a HOVH. To illustrate this transition, we plot in Fig.~\ref{fig:2}(a) the DOS for two representative cases: without HOVH (\(t'=t''=0\), black solid line) and with HOVH (\( t' = 0.10 \), \(t''=0.20\), red solid line). The transition is reflected in the change from the black solid line to the red solid line.  Furthermore, Fig.~\ref{fig:2}(b) shows the noninteracting band dispersion along the high-symmetry path for different \(t''\) values at fixed \(t=1.0, t'=0.20\), providing a direct visualization of the HOVH formation.

In our simulations, we employed a two-dimensional honeycomb lattice with periodic boundary conditions, and set the onsite Coulomb interaction to \(U = 3.0|t|\). This value, which is half the non-interacting bandwidth of the pristine lattice, places the system in a moderately to strongly correlated regime. And the use of \( U = 3.0|t| \) falls within the well-justified range of \( U/|t| = 1-3 \) employed in prior DQMC work by Cheng \textit{et al}.\cite{PhysRevB.91.075410}, which explicitly states the reliability of DQMC simulations within this correlated regime. This standard choice ensures strong correlations are present to investigate the interplay between magnetic fluctuations, superconducting pairing, and the HOVH, even with the inclusion of \( t' \) and \( t'' \). 
The system size was set to \(L =\) 4 and 5, leading to a total of \(N = 2\times3L^2\) lattice sites. This parameter selection enables us to investigate fundamental physical phenomena within a computationally tractable framework while effectively capturing the essential characteristics of the honeycomb lattice structure.  

The choice of QMC methods is particularly suited to this study. Our system resides in an intermediate-to-strong coupling regime (\(U = 3.0|t|\)), where perturbative methods often fail to capture strong electron correlations and fluctuations reliably. QMC provides a numerically exact treatment for such problems. Specifically, DQMC is used for finite-temperature spin susceptibility, as the negative sign problem is not severe, ensuring statistical reliability\cite{PhysRevLett.88.117002,PhysRevD.24.2278,10.1063/1.2839589,PhysRevLett.62.1407} . Concurrently, CPMC is employed as it effectively circumvents the sign problem, enabling extraction of ground-state superconducting pairing correlations\cite{NGUYEN20143344,PhysRevB.55.7464,PhysRevLett.74.3652,PhysRevB.107.245106,PhysRevLett.74.3652,PhysRevB.55.7464,PhysRevB.84.121410,PhysRevB.106.134513}. This combined QMC approach thus offers a rigorous framework for our studies. 

The DQMC methodology employs a stochastic evaluation of the partition function through auxiliary field discretization, where the high-dimensional integration is performed via importance sampling with controlled Trotter errors. In contrast, the CPMC approach projects the ground-state wave function through imaginary-time evolution within a constrained manifold of Slater determinants, enforced by a trial wave function $|\psi_T\rangle$ that maintains positive overlaps throughout the random walk process. Extensive benchmarking confirms the constraint-induced systematic error remains below 3\% for measured observables, with ground-state properties showing remarkable insensitivity to the specific choice of $|\psi_T\rangle$ \cite{PhysRevLett.88.117002}. Our implementation utilizes closed-shell filling configurations and adopts a free-electron trial wave function, $|\psi_T\rangle = \prod_{k \leq k_F} c_k^\dagger |0\rangle$, where the Fermi surface corresponds to the non-interacting system. This construction ensures computational efficiency while preserving physical consistency with the metallic phase under investigation, achieving an optimal balance between numerical tractability and systematic error control.

To study magnetic fluctuations, we defined the spin susceptibility in the \( z \) direction at zero frequency as follows:
\begin{equation}
\chi_s(\mathbf{q}) = \frac{1}{N} \int_0^\beta d\tau \sum_{i,j} e^{i\mathbf{q} \cdot (\mathbf{r}_i - \mathbf{r}_j)} \langle S_i^z(\tau) S_j^z(0) \rangle.
\end{equation}
Here, \( \chi_s(\Gamma) \) quantifies FM correlations, and \( \chi_s(M) \) quantifies AFM correlations.

Within the CPMC framework, we investigate long-range superconducting pairing correlations by analyzing pairing correlation functions for various pairing symmetries. The pairing correlation function is defined as
\begin{equation}
C_\alpha(r) = \frac{1}{N} \sum_{i} \langle \Delta_\alpha^\dagger(\mathbf{r}_i + \mathbf{r}) \Delta_\alpha(\mathbf{r}_i) \rangle.
\end{equation}
For the NN pairing (e.g., \( d+id \) pairing), the pairing operator takes the form\cite{PhysRevB.37.5070,PhysRevB.37.7359}
\begin{equation}
\Delta_{d+id}(\mathbf{r}_i) = \sum_{\delta} \eta_{d+id}(\delta) \left( c_{i\uparrow} c_{i+\delta\downarrow} - c_{i\downarrow} c_{i+\delta\uparrow} \right).
\end{equation}
For the NNN pairing (e.g., \( f_n \) pairing), the pairing operator is given by
\begin{equation}
\Delta_{f_n}(\mathbf{r}_i) = \sum_{\delta'} \eta_{f_n}(\delta') \left( c_{i\uparrow} c_{i+\delta'\downarrow} + c_{i\downarrow} c_{i+\delta'\uparrow} \right).
\end{equation}

The form factors for NN \( d+id \) pairing and NNN \( f_n \) pairing are given as follows:
\begin{equation}
\eta_{d+id}(\delta) = \begin{cases} 
1, & \delta = \vec{\mathbf{a}}_1 \\
e^{i2\pi/3}, & \delta = \vec{\mathbf{a}}_2 ,\\
e^{i4\pi/3}, & \delta = \vec{\mathbf{a}}_3 
\end{cases}
\end{equation}
\begin{equation}
\eta_{f_n}(\delta') = \sin\left( \frac{\mathbf{k} \cdot \delta'}{2} \right), \;  \delta' = \vec{\mathbf{a}}_1, \vec{\mathbf{a}}_2,... \,\vec{\mathbf{a}}_6.
\end{equation}
Their pairing symmetry configurations are sketched in Figs.\ref{Fig:sketch}(c)  and \ref{Fig:sketch}(d), respectively.

\section{Results and Discussion}
\begin{figure}[ht]
\centering
\includegraphics[scale=0.4]{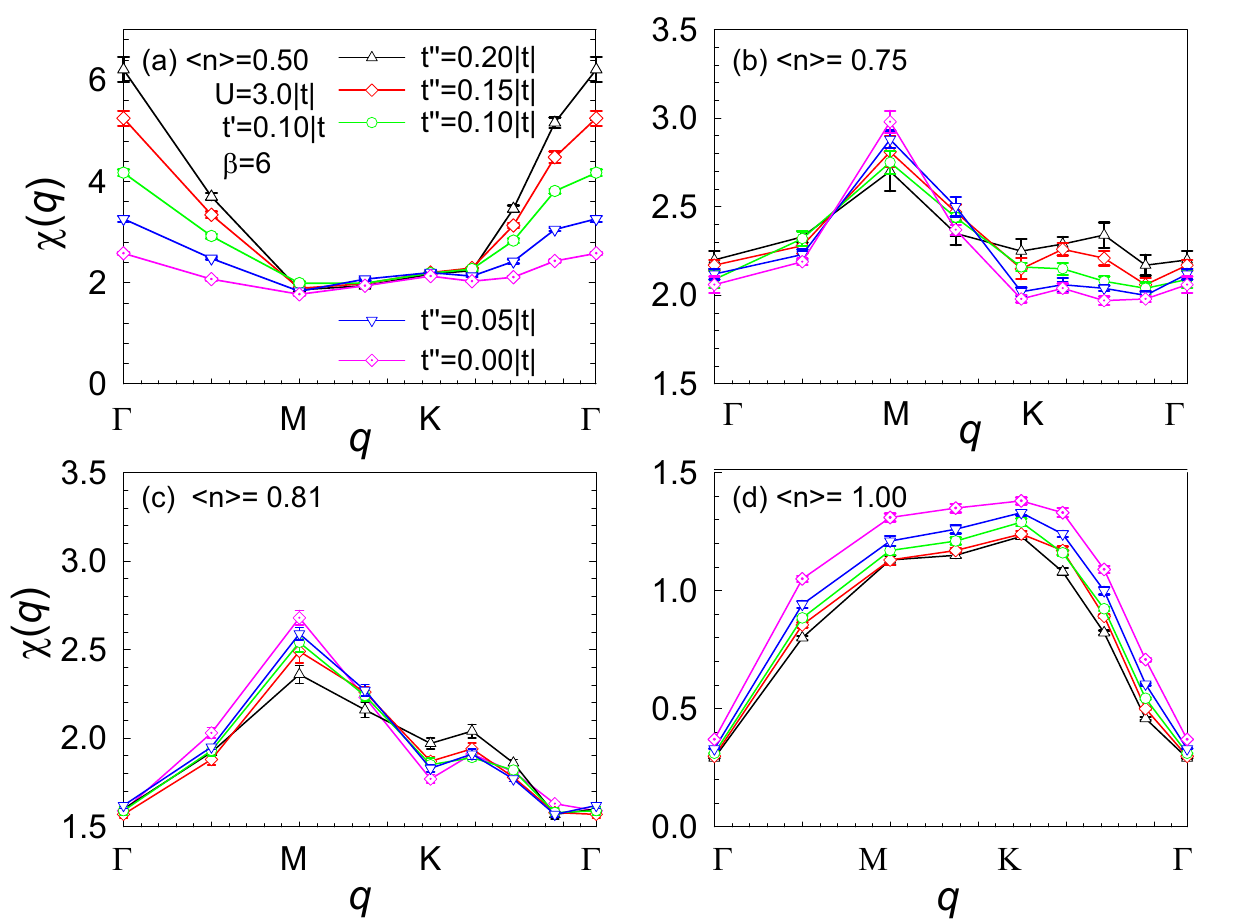}
\caption{The relationship between the spin susceptibility and \( \mathbf{q} \) for different TNN hopping integrals \( t'' \) at fixed \( U = 3.0|t| \) and \( \beta = 6 \) when \(L=4\). (a) \( \langle n \rangle = 0.50 \). (b) \( \langle n \rangle = 0.75 \). (c) \( \langle n \rangle = 0.81 \). (d) \( \langle n \rangle = 1.00 \).}
\label{fig:3}
\end{figure}

We begin by setting the hopping parameters \( t = 1.00 \), \( t' = 0.10 \), and Hubbard interaction \( U = 3.0|t| \), while varying TNN hopping \( t'' \) across different electron fillings \( \langle n \rangle = 0.50 \), 0.75, 0.81, and 1.00. Our study focus on the influence of \( t'' \) on spin susceptibility. The results demonstrate distinct behaviors in spin susceptibility between two filling regions separated by the HOVH singularity. Notably, a crossover from FM to AFM correlations emerges at \( \langle n \rangle = 0.75 \), as shown in Fig.~\ref{fig:3}. To best illustrate the evolution of magnetic fluctuations with \(t''\) within each filling regime, we employed different vertical scales for each subplot. 

Specifically, in the filling region below the HOVH singularity (i.e., \( \langle n \rangle = 0.50\)), the system predominantly exhibits FM behavior, which is significantly enhanced with increasing \( t'' \). Near the HOVH filling at \( \langle n \rangle = 0.75 \), FM and AFM correlations compete: \( t'' \) suppresses fluctuations at the \( M \) point while enhancing it at the \( \Gamma \) points. In contrast, within the filling range from HOVH to half filling (\( \langle n \rangle = 1.0 \)), the system primarily displays AFM correlations that are suppressed by increasing \( t'' \). This momentum-space trend is further corroborated by real-space spin correlation analysis presented in the Appendix B. These results provide further support for previous findings.

\begin{figure}[ht]
\centering
\includegraphics[scale=0.5]{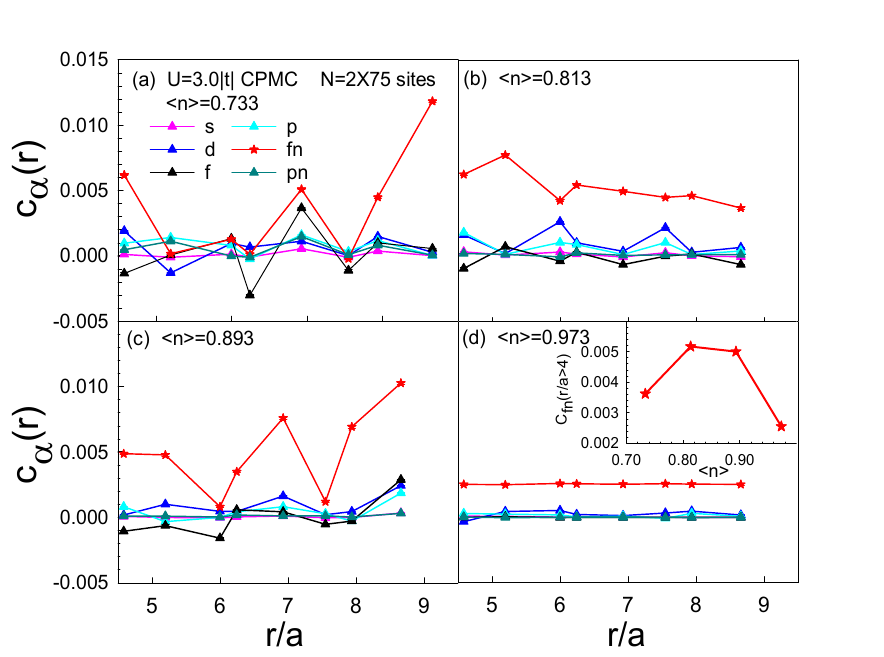}
\caption{The pairing correlation as a function of distance for different pairing symmetries with (a) \( \langle n \rangle = 0.733 \); (b) \( \langle n \rangle = 0.813 \); (c) \( \langle n \rangle = 0.893 \); (d) \( \langle n \rangle = 0.973 \).}
\label{fig:4}
\end{figure}

To explore the property of superconduciting pairing, we set the hopping parameters to \( t = 1.00 \), \( t' = 0.10 \), and \( t'' = 0.20 \), and identify the filling \( \langle n \rangle = 0.881 \) as corresponding to the HOVH singularity. For an on-site Coulomb interaction \( U = 3.0|t| \), we focus on closed-shell fillings near \( \langle n \rangle = 0.881 \).

The results presented in Fig.~\ref{fig:4} reveal that the \( f_n \)-wave pairing symmetry dominates over other symmetries, followed by the \( d+id \) wave. As the electron filling \( \langle n \rangle \) decreases from \( 0.973 \) to \( 0.813 \), the average \( f_n \)-wave pairing correlation is significantly enhanced. In contrast, when \( \langle n \rangle \) deviates further from \( 0.813 \), the \( f_n \)-wave correlations are suppressed. This underscores the pivotal role of the HOVH singularity in enhancing pairing correlations.

\begin{figure}[ht]
\centering
\includegraphics[scale=0.5]{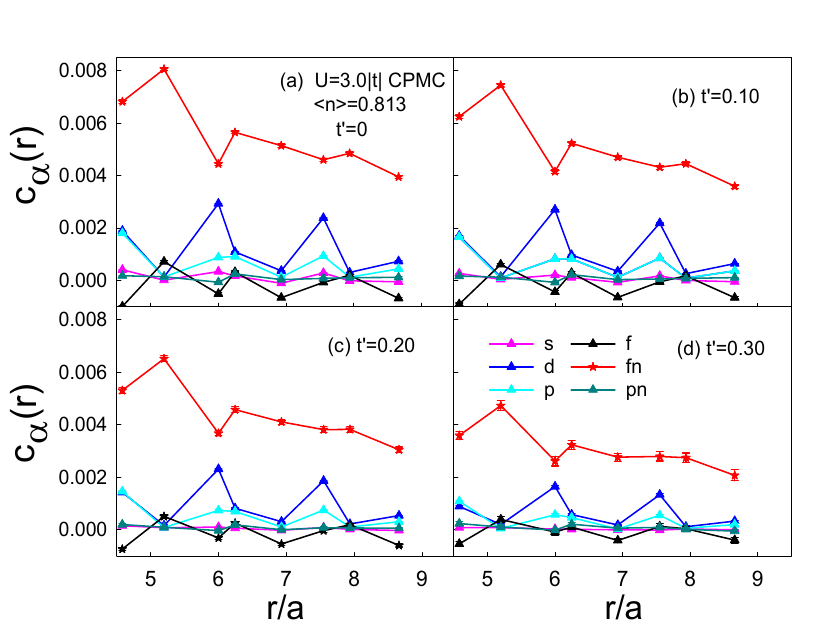}
\caption{Pairing correlation functions for different pairing symmetries with (a) \( t' = 0 \), (b) \( t' = 0.10 \), (c) \( t' = 0.20 \), and (d) \( t' = 0.30 \).}
\label{fig:5}
\end{figure}

To investigate the influence of the NNN hopping \( t' \) on long-range superconducting pairing correlations, we fix the electron filling to \( \langle n \rangle = 0.813 \), with hopping parameters \( t = 1.00 \) and \( t'' = 0.15 \), while systematically varying \( t' \) from 0 to 0.30. As shown in Fig.~\ref{fig:5}, both the \( f_n \) wave and \( d+id \) wave pairing correlation functions exhibit a monotonic suppression as \( t' \) increases.

\begin{figure}[ht]
\centering
\includegraphics[scale=0.5]{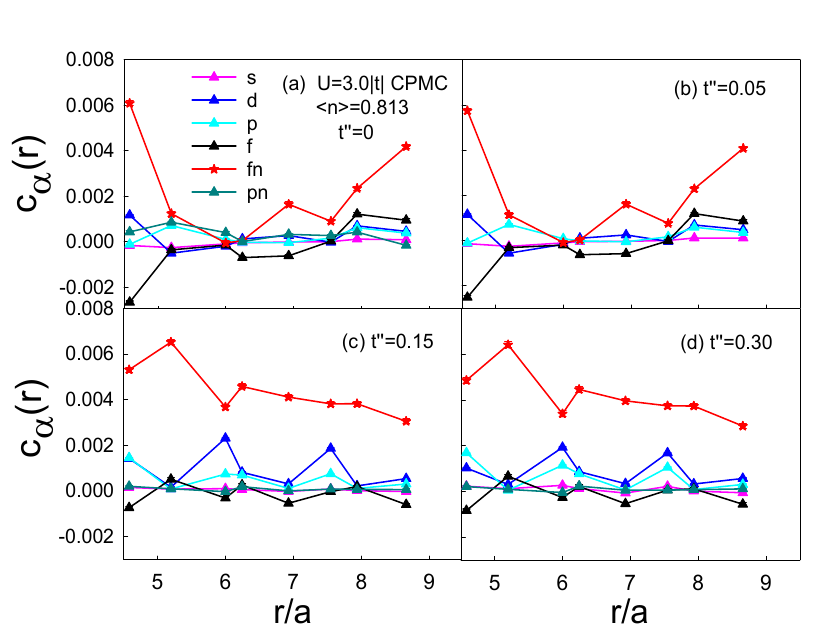}
\caption{Pairing correlation functions for different pairing symmetries with (a) \( t'' = 0 \), (b) \( t'' = 0.05 \), (c) \( t'' = 0.15 \), and (d) \( t'' = 0.30 \).}
\label{fig:6}
\end{figure}

A similar analysis of the effect of \( t'' \) on pairing correlations is performed with \( \langle n \rangle = 0.813 \), \( t = 1.00 \), and \( t' = 0.20 \), while \( t'' \) ranges from 0 to 0.30 in Fig.~\ref{fig:6}. Here, the \( f_n \) wave remains dominant, and at \( t'' = 0.15 \) the superconducting pairing correlations exhibit an unexpected strengthening due to the HOVH effect.
{\color{black}Such a \(t''\) value may be accessible in engineered systems through strain, doping, or substrate interactions\cite{Cao2018B,Levy2010StrainInducedPF}. }

The observed transition from FM to AFM correlations across the HOVH singularity, coupled with the dominant $f_n$-wave superconducting pairing near $\avg{n}$=0.813, reveals a profound interplay between magnetic fluctuations and unconventional superconductivity. Below the HOVH filling $(\avg{n}<0.75$), the enhanced FM correlations with increasing $t''$ suggest a Stoner-type instability that typically competes with spin-singlet pairing. However, the emergence of robust $f_n$-wave superconductivity near $\avg{n}$=0.813, precisely where FM and AFM correlations compete, implies a complex mediation mechanism: while strong FM fluctuations would normally suppress conventional superconductivity, 
the proximity to the HOVH singularity may enable an unconventional pairing channel through the unique momentum-space structure of spin fluctuations near the HOVH point.
These results collectively suggest that the HOVH singularity not only enhances pairing correlations through its singular density of states but also creates a unique magnetic environment where competing FM/AFM fluctuations may actively mediate unconventional superconductivity.

The influence of $t''$ on both FM and AFM correlations, alongside the nonmonotonic enhancement of $f_n$ wave pairing at $t''$=0.15, further suggests that TNN hopping tunes the system through a sweet spot where the HOVH-enhanced DOS cooperates with residual magnetic fluctuations to stabilize $f_n$-wave pairing. This behavior bears resemblance to the spin-fluctuation-mediated pairing scenarios, where competition between different magnetic modes can promote unconventional pairing symmetries. The dominance of $f_n$ wave pairing may be particularly favored by the anisotropic spin fluctuations emerging near the HOVH filling, as evidenced by the $t''$-dependent modulation of susceptibility at $\Gamma$, $K$, and $M$ points.

\begin{figure}[ht]
\centering
\includegraphics[scale=0.5]{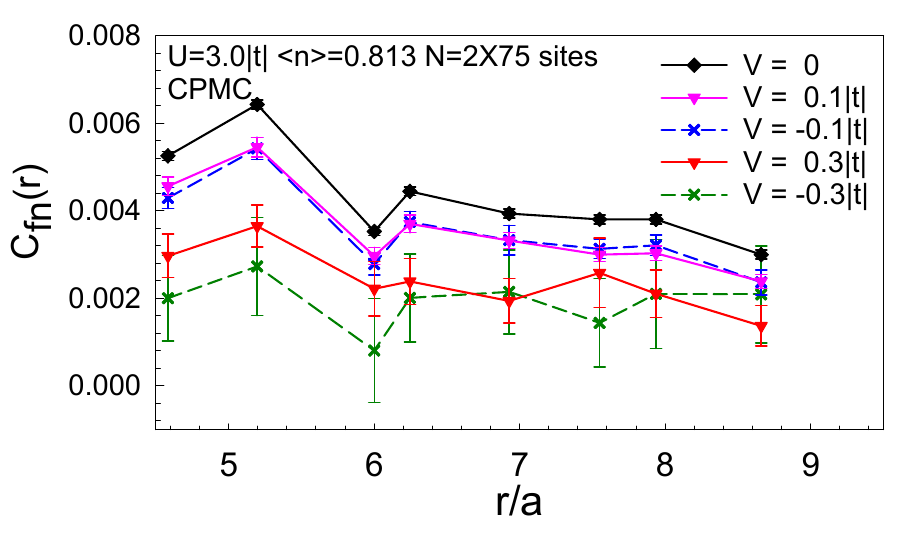}
\caption{ Behavior of pairing correlation functions for \( f_n \)-wave symmetry under varying NN interaction strengths \( V \).}
\label{fig:7}
\end{figure}

Finally, we delve into the specific impact of  \( V \) on \( f_n \)-wave pairing correlations in Fig.~\ref{fig:7}. Increasing \( V \) suppresses \( f_n \)-wave pairing correlations, regardless of its sign, which is similar in many other systems. 
 
\section{Conclusion}
In this work, we have systematically investigated the magnetic correlations and superconducting pairing near HOVH singularities by employing DQMC and CPMC methods. We observed a crossover from FM to AFM correlations in the vicinity of the HOVH filling, driven by the interplay between TNN hopping \(t''\) and electronic DOS divergence. Below the HOVH, \(t''\) amplifies FM fluctuations, while suppressing AFM correlations at higher fillings approaching half filling. In the filling regime \(\langle n \rangle \approx 0.8\), \(f_n\)-wave pairing symmetry dominates, with HOVH-induced power-law divergence in the DOS significantly enhancing superconducting correlations. Interestingly, while larger \(t'\) and \(t''\) values generally suppress \(f_n\)-wave pairing, a critical \(t''=0.15\) triggers an anomalous enhancement of pairing correlations at a fixed $t'$ due to the HOVH effect. These results not only advance our understanding of the competitive mechanisms between magnetic fluctuations and unconventional superconductivity in HOVH systems but also provide a theoretical respect for tailoring correlated electronic phases in graphene-derived quantum materials through strain, doping, or interaction engineering.

\noindent
\section{Acknowledgements}
This work was supported by National Natural Science Foundation of China (Grants
No. 12474218 and No. 12088101) and Beijing Natural Science
Foundation (Grants No. 1242022 and 1252022). The numerical simulations in this work were performed at the HSCC (High-performance Scientific Computing Center) of Beijing Normal University.

\noindent
\section{Data availability}
The data that support the findingsof this article are openly available.\cite{wei_2026_18295499}

\appendix

\setcounter{equation}{0}
\setcounter{figure}{0}
\renewcommand{\theequation}{A\arabic{equation}}
\renewcommand{\thefigure}{A\arabic{figure}}
\renewcommand{\thesubsection}{A\arabic{subsection}}

	\vskip0.3in

\section{Appendix A: FINITE-SIZE ANALYSIS} \label{sec:sfwm}
\setcounter{figure}{7}
\renewcommand{\thefigure}{\arabic{figure}}

{\color{black}To further verify the robustness of our results against finite-size effects, we performed additional large-scale simulations for system size $L=6$.
	
	\vskip0.1in
Fig.~\ref{Fig8} compares the spin susceptibility at $\langle n\rangle=0.81$ between $L=5$ and $6$. The qualitative behavior remains identical across system sizes. This demonstrates the convergence of magnetic correlations with increasing system size.

	\begin{figure}[h]
	\centering
	\adjustbox{trim=0 265 0 72, clip}{%
		\includegraphics[width=0.52\textwidth,height=0.9\textwidth]{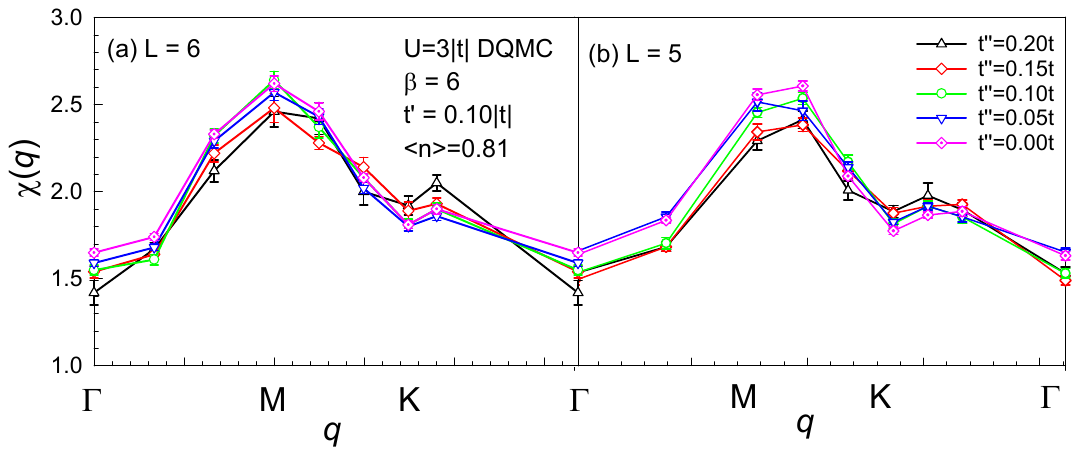}%
	}
	\caption{\label{Fig8} The relationship between the spin susceptibility and \( \mathbf{q} \) for different \( t'' \) at fixed  \( \langle n \rangle = 0.81 \), \( U = 3.0|t| \), and \( \beta = 6 \). (a) \( L=5 \). (b) \( L=6\).}
\end{figure}

	\vskip0.1in

To assess the finite-size effects on the superconducting pairing correlations, we present in Fig.~\ref{Fig9} the pairing correlation functions for different pairing symmetries across different system sizes \(L = 4\) and \(6\). For each size, we select the closed-shell filling closest to the HOVH singularity: \(\langle n \rangle = 0.881\) for \(L=4\), and \(\langle n \rangle = 0.843\) for \(L=6\).

\begin{figure}[h!]
	\includegraphics[width=0.49\textwidth,height=0.23\textwidth]{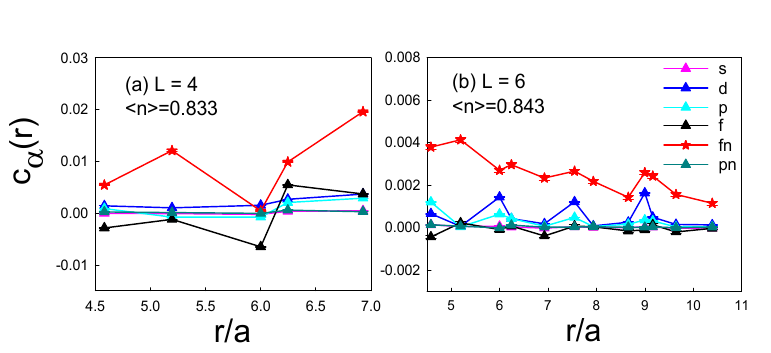}
	\caption{\label{Fig9} Pairing correlation functions for different pairing symmetries at \( t = 1 \), \( t' = 0.20 \), \(t'=0.15\) when (a) \(L=4 \) and (b) \( L=6 \).}
\end{figure}

A direct comparison across  Figs.~\ref{Fig9}(a) and \ref{Fig9}(b) reveals a consistent physical picture: the \(f_n\) wave pairing symmetry dominates over other symmetries near the HOVH filling, regardless of the system size. 
	\vskip0.1in
Furthermore, the anomalous enhancement of \(f_n\) wave pairing at the critical third-nearest-neighbor hopping \(t'' = 0.15\), as discussed in the main text, is also robustly reproduced in the larger \(L=6\) system. This key feature is explicitly demonstrated in Fig.~\ref{Fig10}, which plots the \(f_n\) wave pairing correlation at \(\langle n \rangle = 0.843\) for \(L=6\) under varying \(t''\). 

\begin{figure}[h!]
	\includegraphics[width=0.45\textwidth,height=0.35\textwidth]{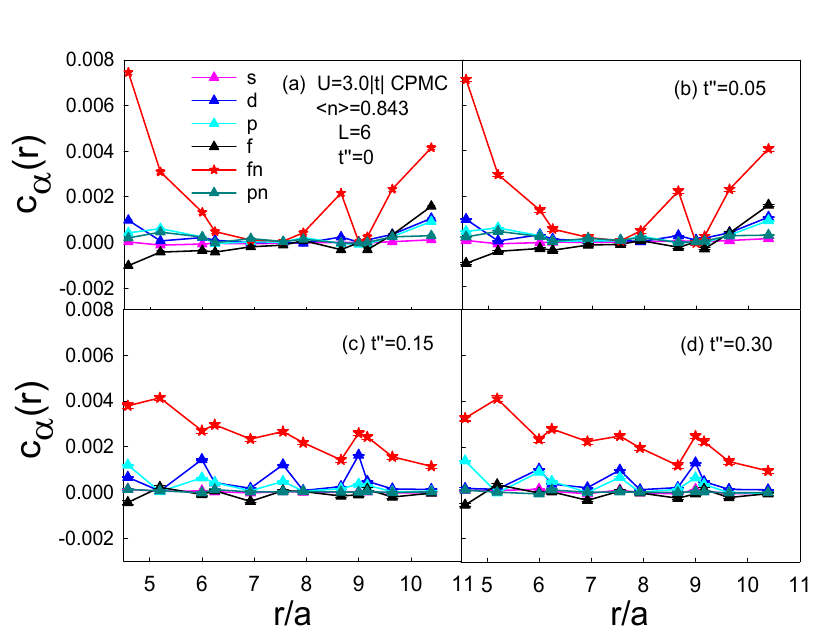}
	\caption{\label{Fig10} Pairing correlation functions for different pairing symmetries at \(L = 6\) when \(t=1, t'=0.20\), and (a) \( t'' = 0 \), (b) \( t'' = 0.05 \), (c) \( t'' = 0.15 \), and (d) \( t'' = 0.30 \).}
\end{figure}

The consistent dominance of $f_n$-wave pairing and the preservation of the critical $t''$ effect across different system sizes confirm that our key findings are intrinsic properties of the system rather than finite-size artifacts. These comprehensive finite-size analyses substantially strengthen the conclusions presented in the main text.
}
\vskip0.1in
\section{Appendix B: REAL-SPACE SPIN CORRELATION ANALYSIS} \label{sec:sfwm}
{\color{black}
To complement the momentum‑space analysis of spin fluctuations presented in Fig.~\ref{fig:3}(d), we examine the equal‑time nearest‑neighbor spin correlation \( S^Z= \langle S_i^z S_j^z \rangle \) for \( U = 3.0|t| \) as \( \langle n \rangle\) approaches 1.00.  

As shown in Fig.~\ref{Fig11}, the AFM correlation (negative \( S_{\langle i,j\rangle} \)) is systematically suppressed as \( t'' \) increases, while changes in \( t' \) show negligible effect at \(<n>=1.00\). This real‑space result provides direct verification of the corresponding trend observed in momentum space in Fig.~\ref{fig:3}(d): the reduction in \( \chi_s(M) \) with increasing \( t'' \). 

\begin{figure}[h!]
	\includegraphics[width=0.43\textwidth,height=0.32\textwidth]{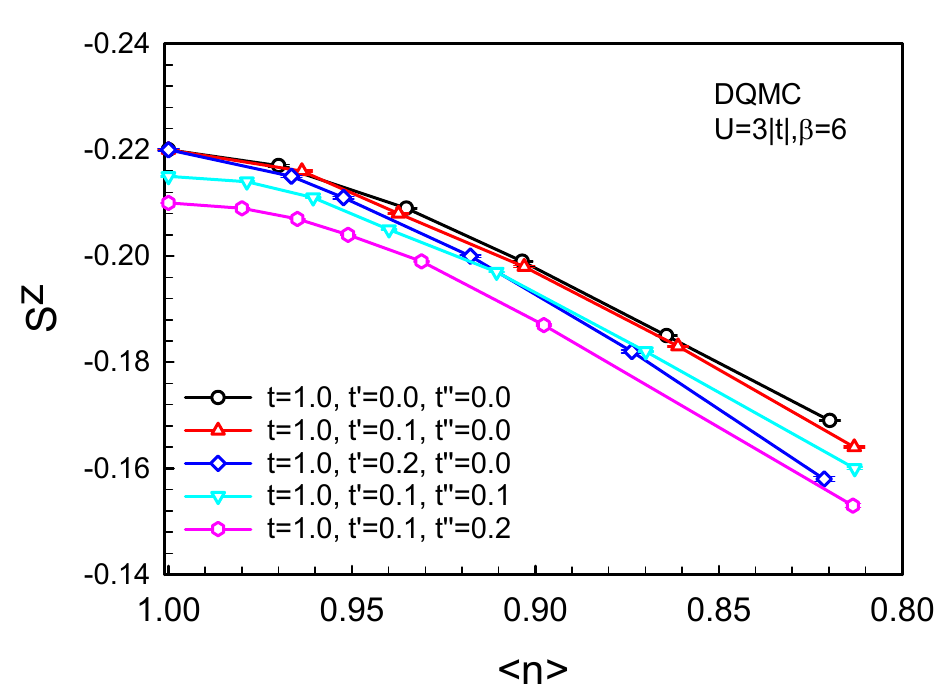}
	\caption{\label{Fig11} The evolution of the equal‑time nearest‑neighbor spin correlation with filling for different \(t'\) and \(t''\)  with $U=3.0|t|$ and $L=4$.}
\end{figure}

}
\bibliography{HVHSReferences}
\end{document}